\documentclass[11pt,epsf]{article}

\usepackage{epsfig}
\usepackage{amssymb}
\usepackage{graphicx}
\usepackage{color}
\usepackage{subfigure}

\def\be{\begin{equation}}\def\ba{\begin{eqnarray}}
\def\ee{\end{equation}}\def\ea{\end{eqnarray}}

\makeatletter

\@addtoreset{equation}{section} \makeatother
\def\be{\begin{equation}}
\def\ee{\end{equation}}
\def\ba{\begin{array}}
\def\ea{\end{array}}
\def\bea{\begin{eqnarray}}
\def\eea{\end{eqnarray}}
\def\nn{\nonumber}

\def\eq#1{(\ref{#1})}

\def\m{\mu}

\def\r{\rho}
\def\s{\sigma}

\def\det{{\rm det}}

\def\sss{\scriptscriptstyle}
\def\nn{\nonumber}

\begin{document}

\begin{titlepage}
\title{\vskip -60pt
\vskip 20pt Holographic Meson Spectra in the Dense Medium with
Chiral Condensate}
\author{
Chanyong Park$^a$\footnote{e-mail : cyong21@sogang.ac.kr}, Bum-Hoon
Lee$^{ab}$\footnote{e-mail : bhl@sogang.ac.kr }, Sunyoung
Shin$^c$\footnote{e-mail : shin@theor.jinr.ru}}
\date{}
\maketitle \vspace{-1.0cm}
\begin{center}
~~~
\it $^a\,$Center for Quantum Spacetime, Sogang University, Seoul 121-742, Korea\\
\it $^b\,$Department of Physics, Sogang University, Seoul 121-741,Korea \\
\it $^c\,$Bogoliubov Laboratory of Theoretical Physics, JINR, 141980
Dubna, Moscow region, Russia
~~~\\
~~~\\
\end{center}

\begin{abstract}
We study two $1/N_c$ effects on the meson spectra by using the
AdS/CFT correspondence where the $1/N_c$ corrections from the chiral
condensate and the quark density are controlled by the gravitational
backreaction of the massive scalar field and $U(1)$ gauge field
respectively. The dual geometries with zero and nonzero current
quark masses are obtained numerically. We discuss meson spectra and
binding energy of heavy quarkonium with the subleading corrections
in the hard wall model.
\end{abstract}

{\small
\begin{flushleft}
\end{flushleft}}
\end{titlepage}
\newpage

\tableofcontents 

\section{Introduction}
\setcounter{equation}{0}
Holographic QCD provides descriptions of the strongly interacting
regime of the gauge theory based on the anti-de Sitter/conformal
field theory (AdS/CFT) correspondence
\cite{Maldacena:1997re,Gubser:1998bc,Witten:1998qj}. The properties
of the gauge theory including confinement, chiral symmetry breaking,
supersymmetry reduction and glueballs are discussed in
\cite{Witten:1998zw,Polchinski:2000uf,Klebanov:2000hb,Maldacena:2000yy,Polchinski:2001tt,BoschiFilho:2002vd,
Kruczenski:2003uq}. QCD-like models are proposed in
\cite{Sakai:2004cn,Sakai:2005yt,Erlich:2005qh,Da
Rold:2005zs,Brodsky:2003px,Evans:2005ip,Karch:2006pv,Csaki:2006ji,Gursoy:2007cb,deTeramond:2005su}.
Baryons and chemical potential are considered in
\cite{Sakai:2004cn,deTeramond:2005su,Hong:2006ta,Kim:2006gp,Horigome:2006xu,Parnachev:2006ev,
Nakamura:2006xk,Kobayashi:2006sb,Domokos:2007kt,Sin:2007ze,Kim:2007zm,Nakamura:2007nx}.

One of the simplest bottom-up approach is the hard wall model
\cite{Erlich:2005qh,Da Rold:2005zs} where the confinement is
realized by introducing an infrared cut-off in the AdS spacetime.
The predictions of the hard wall model match measured values of
meson masses within 10\% error.

AdS geometry is dual to the large $N_c$ limit of $U(N_c)$ gauge
theory on the boundary, which is conformal. Pure AdS geometry can be
considered as a UV fixed point of the holographic QCD. If we
consider a subleading correction having a typical scale, the
conformal symmetry should be broken. As a result, the dual geometry
should be modified to include the gravitational backreaction of the
bulk field dual to the subleading correction, which may improve the
hard wall model.

There are two important holographic subleading corrections. One is a
correction from the hadronic medium and the other is a correction
from the chiral condensate. The subleading correction from the
hadronic medium is the gravitational backreaction of local $U(1)$
gauge fields. The Reissner-Nordstr\"{o}m AdS black hole (RN AdS BH)
and thermal charged AdS (tcAdS), which is the zero mass limit of the
RN AdS BH, are proposed as the corresponding geometries for
deconfinement and confinement phases respectively
\cite{Sin:2007ze,Lee:2009bya,Park:2011zp}. The Hawking-Page
transition between the geometries, and the meson spectra and the
binding energy of heavy quarkonium on the tcAdS are discussed in
\cite{Sin:2007ze,Lee:2009bya,Jo:2009xr,Park:2009nb}. The subleading
correction from the chiral condensate is the gravitational
backreaction of a massive scalar field
\cite{Kim:2007em,Shock:2006gt,Wu:2007tza,Lee:2010dh}. In
\cite{Kim:2007em} the contribution of the scalar field to the
Hawking-Page transition is investigated. In
\cite{Shock:2006gt,Wu:2007tza}, holographic models capturing the
gravitational backreaction of the scalar field are constructed. In
\cite{Lee:2010dh}, the dual geometries with zero and nonzero quark
masses are obtained by numerically solving the equations of motion
of the bulk action including the scalar field. The light meson
spectra and the binding energy of heavy quarkonium are discussed on
both backgrounds.

In this paper we study subleading corrections from the chiral
condensate and the hadronic medium as an extension of
\cite{Lee:2009bya,Jo:2009xr,Park:2009nb,Lee:2010dh}. We numerically
solve the equations of motion of the bulk action including a massive
scalar field and $U(1)$ gauge fields to obtain the dual geometries
with zero and nonzero current quark masses. In QCD phenomenology, it
is known that the chiral condensate gets reduced in the hadronic
medium \cite{Drukarev:1988kd}. We constrain the dual geometries with
a model-independent relation between the chiral condensate and the
quark density. We calculate meson masses and binding energy of heavy
quarkonium on the geometrical backgrounds. By comparing the results
with \cite{Lee:2009bya,Jo:2009xr,Park:2009nb,Lee:2010dh}, we discuss
the effects of the chiral condensate and the quark density.

Organization of this paper is as follows. In section 2, the bulk
action and the equations of motion of it for the dual geometries are
discussed. In section 3 and 4, the equations of motion of the bulk
action are solved with zero and nonzero current quark masses. The
meson masses and the binding energy of heavy quarkonium are obtained
on the asymptotic AdS geometries. In section 5, we summarize the
results.

\section{Asymptotic AdS background}
\setcounter{equation}{0}
The gravity action in the bulk is
\begin{eqnarray}
\mathcal{S}&=&\int
d^5x\sqrt{-G}\left\{\frac{1}{2\kappa^2}\left(\mathcal{R}-2\Lambda\right)\right.\nn\\
&&\left.\hspace{2cm}-\mathrm{Tr}\left[|D\Phi|^2+m^2|\Phi|^2
+\frac{1}{4g^2}\left(F^{(L)}_{MN}F^{(L)MN}+F^{(R)}_{MN}F^{(R)MN}\right)\right]\right\},\label{eq:act5d}
\end{eqnarray}
where $m^2=-\frac{3}{R^2}$ and $\Lambda=-\frac{6}{R^2}$. We follow
the convention that $g^2=\frac{12\pi^2R}{N_c}$ and
$\kappa^2=\frac{\pi^2R^3}{4N_c^2}$ \cite{Erlich:2005qh}. We consider
the case of $N_f=2$ and $N_c=3$. The superscripts $(L)$ and $(R)$
are of $SU(N_f)_L\times SU(N_f)_R$ flavor symmetry with
$F^{(L,R)}_{MN}=\partial_MA_N^{(L,R)}-\partial_NA_M^{(L,R)}-i\Big[A_M^{(L,R)},A_N^{(L,R)}\Big]$.
The covariant derivative of the complex scalar field is defined by
$D_M\Phi=\partial_M\Phi-iA_M^{(L)}\Phi+i\Phi A_M^{(R)}$. We set
\begin{eqnarray}
\Phi(z)=\frac{1}{2\sqrt{N_f}}\phi(z)\mathbf{1}_fe^{i2\pi^a(z)T^a} ,
\label{eq:defphi}
\end{eqnarray}
where the modulus of the complex scalar field is regarded as a
background field giving the gravitational backreaction while $\pi^a
(z)$ corresponding to the scalar meson is regarded as fluctuations.

The background geometries of our interest are obtained from a
gravity action with the massive scalar field of the Lagrangian
(\ref{eq:act5d}) and $U(1)$ gauge interaction. The corresponding
action is
\begin{eqnarray}
{\mathcal{S}}=\int
d^5x\sqrt{-G}\Big\{\frac{1}{2\kappa^2}({\mathcal{R}}-2\Lambda)
-\frac{1}{4}\Big[\left(\partial_M\phi\right)^2+m^2\phi^2\Big]-\frac{1}{4g^2}F_{MN}F^{MN}\Big\},\label{eq:actbg}
\end{eqnarray}
where $\phi(z)$ is the scalar field in (\ref{eq:defphi}). We choose
an ansatz for the asymptotic AdS metric in the Fefferman-Graham
coordinate as
\begin{eqnarray}
ds^2=\frac{R^2}{z^2}\left[-F(z)dt^2+G(z)dx^2+dz^2\right],\label{eq:metricansatz}
\end{eqnarray}
where $R$ is the AdS radius. We take $R=1$. For the AdS black hole
without the chiral condensate and the quark density, the components
of the metric (\ref{eq:metricansatz}) are given by
\begin{eqnarray}
F(z)&=&\frac{(1-Mz^4)^2}{1+Mz^4},\nonumber\\
G(z)&=&1+Mz^4, \label{eq:fg}
\end{eqnarray}
where $M$ is the black hole mass, which is asymptotically the AdS
space. In the AdS spacetime, the modulus of the scalar field
$\phi(z)$ is
\begin{eqnarray}
\phi(z)=m_qz+\sigma z^3,\label{eq:scalar}
\end{eqnarray}
where $m_q$ and $\sigma=\langle\bar{q}q\rangle$ are related to the
current quark mass and the chiral condensate of QCD. The $U(1)$
gauge fields in (\ref{eq:actbg}) are $A_0=A_0(z)$ whereas $A_M=0$
for $M=1,2,3,z$. In the RN AdS BH or tcAdS, the time component
vector field is
\begin{eqnarray}
A_0(z)=\mu-Qz^2,\label{eq:vector}
\end{eqnarray}
where $\mu$ and $Q$ are related to the chemical potential and the
quark number density.

Now, we generalize the metric (\ref{eq:metricansatz}) including the
effects of the chiral condensate and the quark density. The Einstein
equations and the equations of motion for $\phi(z)$ and $A_0(z)$ are
\begin{eqnarray}
&&0=\frac{1}{4}\left\{-\frac{2z^2\kappa^2A_0^{\prime2}}{g^2}+
\frac{F}{z^2G}\Big[\kappa^2G(3\phi^2-z^2\phi^{\prime2})
-6z(-3G^\prime+zG^{\prime\prime})\Big]\right\},\nonumber\\
&&0=\frac{1}{4}\Big\{\kappa^2G\left(-\frac{3\phi^2}{z^2}+\phi^{\prime2}
-\frac{2z^2A_0^{\prime2}}{g^2F}\right)+G^\prime\left(-\frac{12}{z}+\frac{2F^\prime}{F}\right)
-\frac{G^{\prime2}}{G}\nonumber\\
&&\hspace{1.5cm}-G\left(\frac{F^{\prime2}}{F^2}+\frac{6F^\prime}{zF}-\frac{2F^{\prime\prime}}{F}\right)
+4G^{\prime\prime}\Big\},\nonumber\\
&&0=\frac{1}{4}\Big\{\kappa^2\left(-\frac{3\phi^2}{z^2}-\phi^{\prime2}+\frac{2z^2A_0^{\prime2}}{g^2F}\right)
+\frac{3G^\prime}{G}\left(-\frac{6}{z}+\frac{G^\prime}{G}\right)
+\frac{3F^\prime}{F}\left(-\frac{2}{z}+\frac{G^\prime}{G}\right)\Big\},\nonumber\\
&&0=\frac{G^{1/2}}{2z^2F^{3/2}}\Big\{A_0^\prime\left[zGF^\prime+F(2G-3zG^\prime)\right]
-2zFGA_0^{\prime\prime}\Big\},\nonumber\\
&&0=3\phi-3z\phi^\prime+\frac{1}{2}z^2\phi^\prime\frac{F^\prime}{F}
+\frac{3}{2}z^2\phi^\prime\frac{G^\prime}{G}+z^2\phi^{\prime\prime}.
\label{eq:geoeom}
\end{eqnarray}
We solve the equations to obtain the geometries with zero and
nonzero quark masses, and calculate the meson masses and the binding
energy of heavy quarkonium on each geometry.

\section{Mesons with a zero $m_q$}
\setcounter{equation}{0}
\subsection{Light meson spectra}\label{sec:masszeromq}
We solve (\ref{eq:geoeom}) with a zero quark mass, $m_q=0$, which
corresponds to the chiral limit showing the spontaneous symmetry
breaking effect caused by the chiral condensation. In the UV limit,
the scalar field $\phi(z)$ and the vector field $A_0(z)$ should be
(\ref{eq:scalar}) and (\ref{eq:vector}) respectively. We also use
the fact that in the absence of the chiral condensate and the quark
density, the solution should be (\ref{eq:fg}). The perturbative
solutions to (\ref{eq:geoeom}) near the boundary, up to
${\mathcal{O}}(z^{12})$ are
\begin{eqnarray}
F(z)&=&1-3Mz^4+\frac{20Q^2\kappa^2-3g^2\kappa^2\sigma^2}{36g^2}z^6+4M^2z^8
+\frac{M\kappa^2(-104Q^2+3g^2\sigma^2)}{60g^2}z^{10},\nonumber\\
G(z)&=&1+Mz^4+\frac{-4Q^2\kappa^2-3g^2\kappa^2\sigma^2}{36g^2}z^6
+\frac{M\kappa^2(8Q^2-3g^2\sigma^2)}{180g^2}z^{10}, \nonumber\\
\phi(z)&=&\sigma
z^3+\frac{\kappa^2\sigma(-2Q^2+3g^2\sigma^2)}{48g^2}z^9+\frac{3M^2\sigma}{10}z^{11},\nonumber\\
A_0(z)&=&\mu-Qz^2+MQz^6+\frac{-16Q^3\kappa^2-3Qg^2\kappa^2\sigma^2}{144g^2}z^8-M^2Qz^{10}.
\label{eq:bgdzeromq}
\end{eqnarray}
We set $M=0$, as we are interested in the meson spectra in the
confining phase. From the above perturbative solutions, we can
easily find the full numerical solutions depending on parameters,
$\s$ and $Q$. Note that if we concentrate on the canonical ensemble
$\m$ is not important for investigating the meson spectra.

The perturbation of the massive scalar field and the bulk gauge
fields describe the pion, vector meson and axial-vector meson
respectively. We transform perturbed gauge fields
$a^{\sss{(L,R)}}_{\sss{M}}$ of $SU(N_f)_L\times SU(N_f)_R$ flavor
symmetry to vector and axial-vector fields:
\begin{eqnarray}
v_{\sss{M}}=\frac{1}{2}(a^{\sss{(L)}}_{\sss{M}}+a^{\sss{(R)}}_{\sss{M}}),
~~~a_{\sss{M}}=\frac{1}{2}(a^{\sss{(L)}}_{\sss{M}}-a^{\sss{(R)}}_{\sss{M}}).
\end{eqnarray}
The action describing the mesons is
\begin{eqnarray}
{\Delta\mathcal{S}}=\int
d^5x\sqrt{-G}\left[-\frac{1}{2}(\phi\partial_{\sss{M}}\pi-a_{\sss{M}}\phi)
(\phi\partial^{\sss{M}}\pi-a^{\sss{M}}\phi)-\frac{1}{4g^2}(f_{\sss{MN}}^{\sss{(V)}}f^{\sss{MN(V)}}
+f_{\sss{MN}}^{\sss{(A)}}f^{\sss{MN(A)}})\right],
\end{eqnarray}
where the $\Phi$ is as defined in (\ref{eq:defphi}) and
$f^{\sss{(V)}}_{\sss{MN}}$ and $f^{\sss{(A)}}_{\sss{MN}}$ are the
field strengths of the gauge fields $v_{\sss{M}}$ and $a_{\sss{M}}$.
The Lagrangian is in the gauge where $v_z=0$ and $a_z=0$. The axial
vector can be decomposed into a transverse component and a
longitudinal component:
\begin{eqnarray}
a_\mu=\bar{a}_\mu+\partial_\mu\chi.
\end{eqnarray}
We set $\bar{a}_{\sss{0}}=0$ as the Lorentz boost symmetry is not
manifest. The equations of motion for $v_i$, $a_i$, $\pi$ and $\chi$
are
\begin{eqnarray}
&&\partial_z\left(\frac{1}{z}F^{\frac{1}{2}}G^{\frac{1}{2}}\partial_zv_i\right)
+\frac{1}{z}F^{-\frac{1}{2}}G^{\frac{1}{2}}m_{\sss{V}}^2v_i=0,\nonumber\\
&&\partial_z\left(\frac{1}{z}F^{\frac{1}{2}}G^{\frac{1}{2}}\partial_z\bar{a}_i\right)
-\frac{1}{z^3}F^{\frac{1}{2}}G^{\frac{1}{2}}\left(g^2\phi^2-\frac{z^2}{F}m_{\sss{A}}^2\right)\bar{a}_i=0,\nonumber\\
&&\partial_z\left(\frac{1}{z}F^{-\frac{1}{2}}G^{\frac{3}{2}}\partial_z\chi\right)
+g^2\phi^2\frac{1}{z^3}F^{-\frac{1}{2}}G^{\frac{3}{2}}(\pi-\chi)=0,\nonumber\\
&&m_\pi^2\partial_z\chi-g^2\frac{F}{z^2}\phi^2\partial_z\pi=0,
\label{eq:mesonmass}
\end{eqnarray}
where $m_{\sss{V}}$, $m_{\sss{A}}$ and $m_\pi$ are the masses of
vector meson, axial-vector meson and pion respectively. We solve
(\ref{eq:mesonmass}) to obtain the meson masses in the background
geometry numerically obtained from (\ref{eq:bgdzeromq}).

The dual geometry corresponding to the ground state contains the
chiral condensate as well as the quark density. The effects of the
quark density and the chiral condensate on the meson spectra are
discussed separately in \cite{Jo:2009xr,Lee:2010dh}. As the quark
density increases, the $\r$-meson mass and the $a_1$-meson mass
increase \cite{Jo:2009xr}. As the chiral condensate increases, the
$\r$-meson mass decreases whereas the $a_1$-meson mass increases
\cite{Lee:2010dh}. Thus it is worth investigating the effect of both
$1/N_c$ corrections to the meson spectra and which of them is more
dominant. In phenomenology, the chiral condensate and the quark
density are not independent. One of the known relations between them
is
\begin{eqnarray}
\frac{\sigma}{\sigma_0}=1-0.35\frac{\rho_B}{\rho_0},\label{eq:con_den}
\end{eqnarray}
where $\r_B$ is the baryon number density and related to the quark
number density with $Q/N_c$. The relation is considered to be
model-independent, on the assumption that the pion-nucleon sigma
term $\Sigma_{\pi N}\simeq45\mathrm{MeV} $ \cite{Drukarev:1988kd}.
The quark density reduces the chiral condensate and it leads to the
modification of the medium requiring non-trivial boundary conditions
in the gravity set-up. Instead of finding such boundary conditions,
we impose the relation (\ref{eq:con_den}) on the geometry
(\ref{eq:bgdzeromq}).

\begin{figure}[h!]
\begin{center}
\epsfxsize=7cm
   \epsfbox{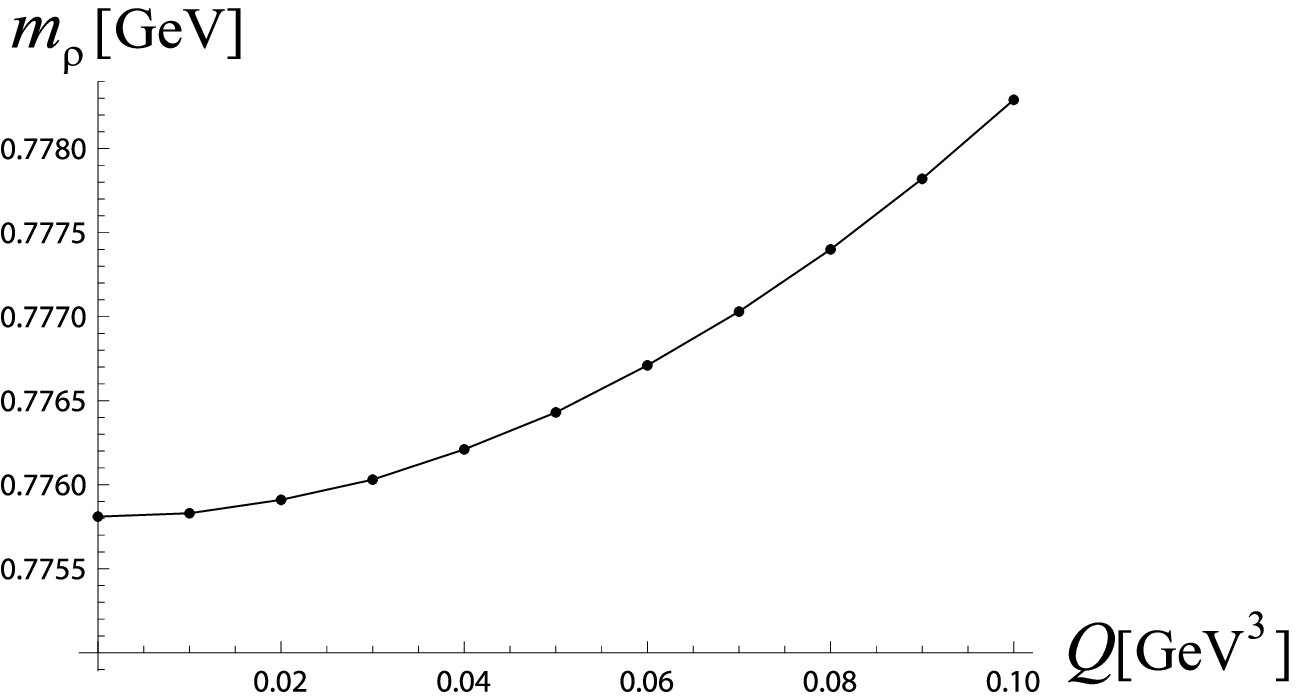}\\
   (a)\\~\\
$\begin{array}{cc}
  \epsfxsize=7cm
   \epsfbox{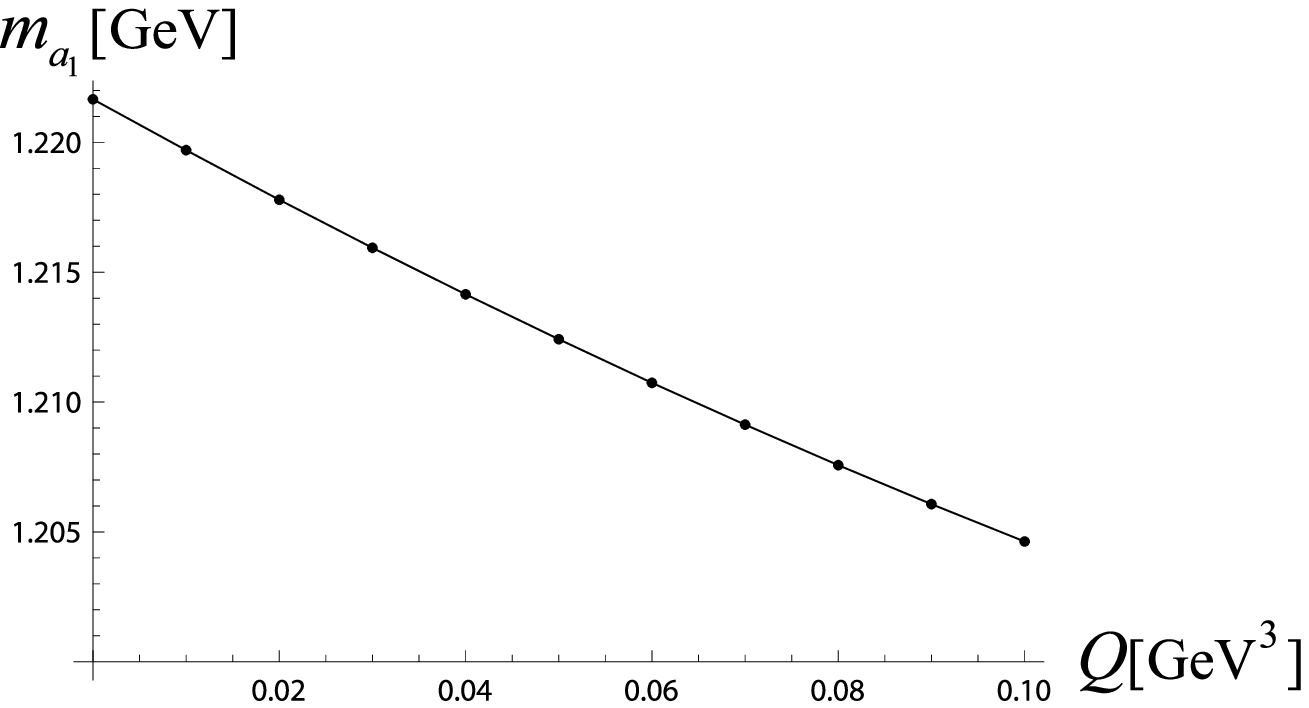}
  &
  \epsfxsize=7cm
   \epsfbox{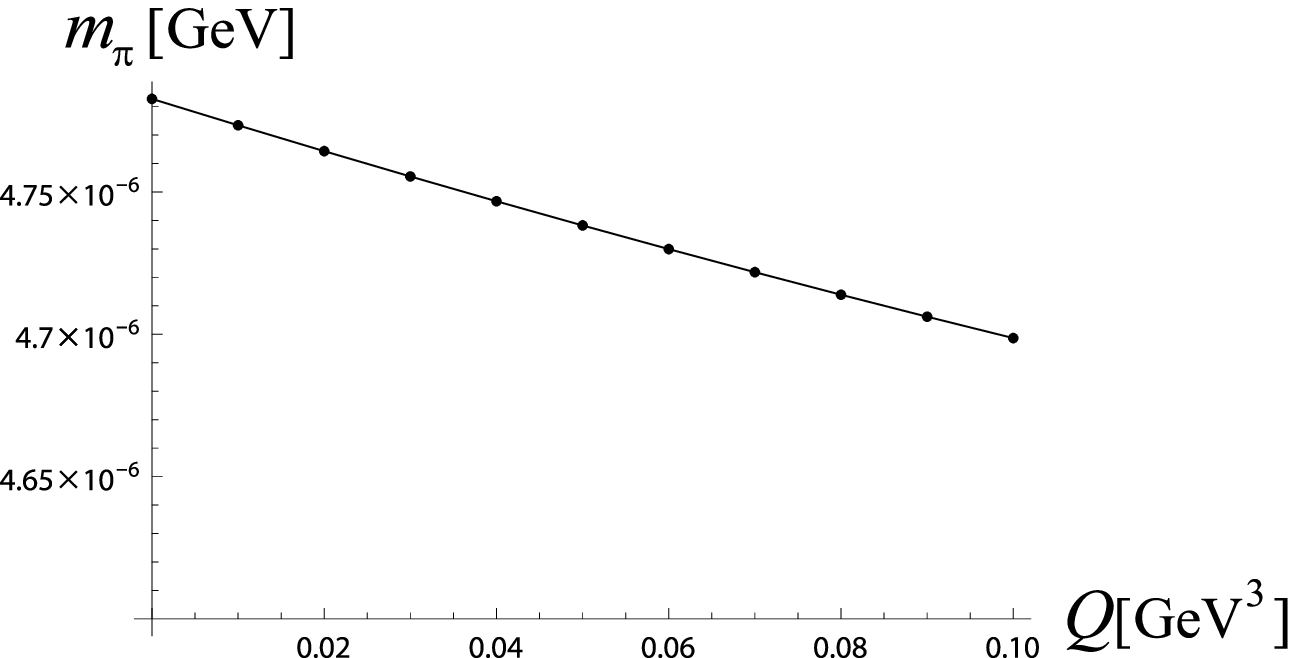}\\
   \mathrm{(b)}&\mathrm{(c)}
\end{array}$
  \caption{ Meson masses with $m_q=0$. (a)$\r$-meson
   (b)$a_1$-meson
   (c)pion}\label{fig:mass1}
 \end{center}
\end{figure}

\begin{figure}[h!]
\begin{center}
$\begin{array}{cc} \epsfxsize=7cm \epsfbox{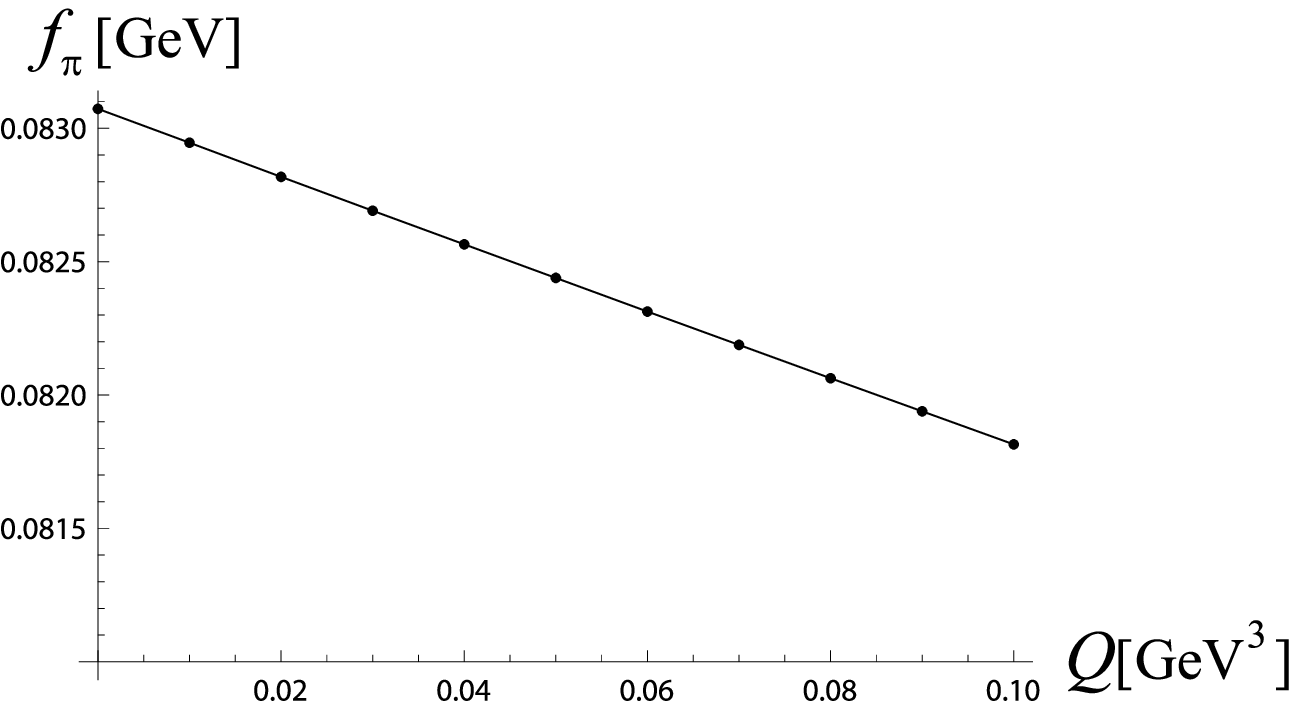} &\epsfxsize=7cm
\epsfbox{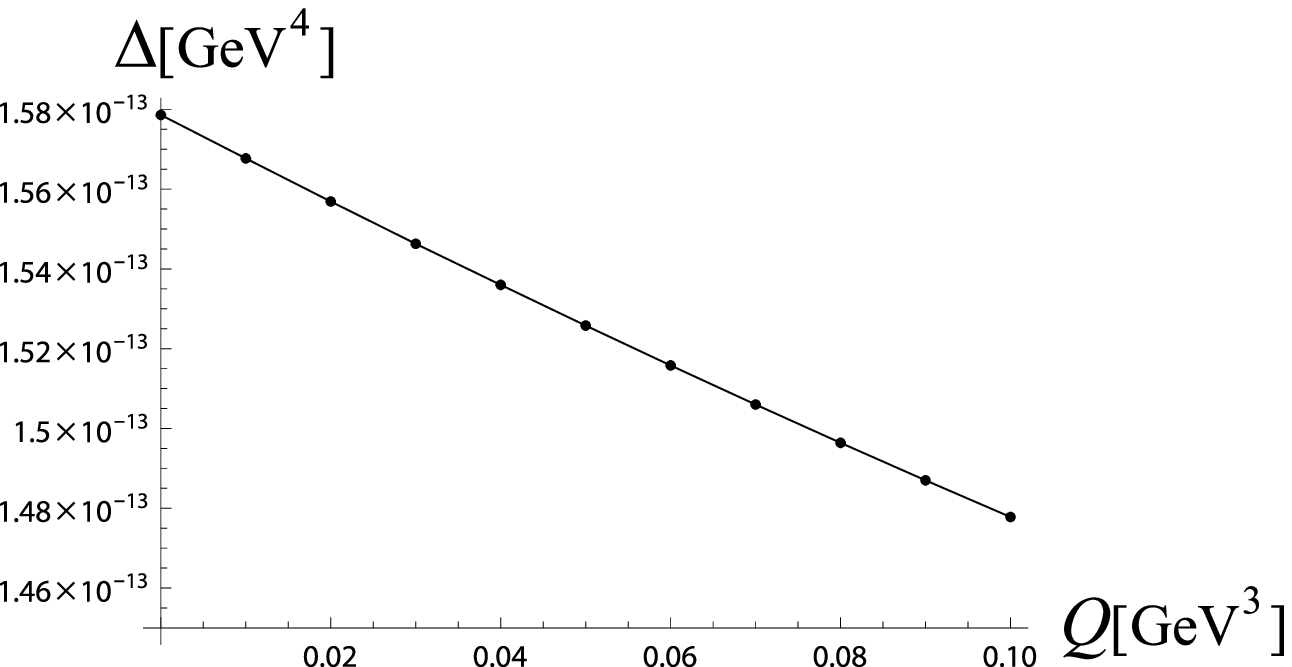}\\
\mathrm{(a)}&\mathrm{(b)}
\end{array}$
\caption{(a)pion decay constant (b)deviation from GOR relation.
$m_q=0$.}\label{fig:decay_gor1}
\end{center}
\end{figure}

The quark density becomes a parameter. In our notation,
$Q/Q_0=\rho_B/\rho_0$ with $Q_0=1\mathrm{GeV^3}$ as a unit. We also
choose
\begin{eqnarray}
\sigma_0=(0.304\mathrm{GeV})^3,~~z_{IR}=1/(0.3227\mathrm{GeV}),\label{eq:sig_ir}
\end{eqnarray}
of which the physical aspects are discussed in \cite{Lee:2010dh}. We
impose Dirichlet and Neumann boundary conditions at $z=0$ and
$z=z_{IR}$ respectively. The pion decay constant $f_\pi$ is
\begin{eqnarray}
f_\pi^2=-\left.\frac{1}{g^2}\frac{\partial_z\bar{a}(0)}{z}\right|_{z=0},
\end{eqnarray}
where $\bar{a}(0)$ is a solution of the second equation of
(\ref{eq:mesonmass}) with $m_{\sss{A}}=0$, satisfying the boundary
conditions $\bar{a}(0)=1$ and $\partial_z\bar{a}(z_{IR})=0$
\cite{Erlich:2005qh}. The result is shown in Table \ref{tab:mass1},
Figure \ref{fig:mass1} and Figure \ref{fig:decay_gor1}.

The $\r$-meson mass increases as the quark density increases. This
is qualitatively consistent with the meson spectra depending on the
chiral condensate \cite{Lee:2010dh} and the quark density
\cite{Jo:2009xr}. The $a_1$-meson mass and pion mass, however,
decrease with the quark density. This is consistent with the mass
spectra depending on the chiral condensate \cite{Lee:2010dh}, but
contrary to the spectra depending on the quark density
\cite{Jo:2009xr}. It shows that the effect of the chiral condensate
dominates the effect of the quark density.

The pion decay constant decreases with the quark density. From
Gell-Mann--Oakes--Renner (GOR) relation
\begin{eqnarray}
f_\pi^2m_\pi^2=2m_q\sigma, \label{eq:gor}
\end{eqnarray}
the deviation is $\Delta=f_\pi^2m_\pi^2$ since $m_q=0$. The GOR
relation is satisfied up to $10^{-13}\mathrm{GeV^4}$.
\begin{table}[h!]
\begin{center}
\begin{tabular}{|c|c|c|c|c|c|}
\hline
$Q(\mathrm{GeV}^3)$ &$m_{\rho}(\mathrm{GeV})$ & $m_{a_1}(\mathrm{GeV})$ & $m_\pi(\mathrm{GeV})$ & $f_\pi(\mathrm{GeV})$ & $\Delta(\mathrm{GeV}^4)  $         \\
\hline \hline
$0.00$              & $0.77581$     & $1.22166$      & $4.78271\times 10^{-6}$ &$8.3073\times10^{-2}$& $1.5786\times10^{-13}$\\
\hline
$0.01$              & $0.77583$     & $1.21970$      & $4.77344\times 10^{-6}$ &$8.2946\times10^{-2}$& $1.5677\times10^{-13}$\\
\hline
$0.02$              & $0.77591$     & $1.21779$      & $4.76435\times 10^{-6}$ &$8.2818\times10^{-2}$& $1.5569\times10^{-13}$\\
\hline
$0.03$              & $0.77603$     & $1.21594$      & $4.75546\times 10^{-6}$ &$8.2691\times10^{-2}$& $1.5463\times10^{-13}$\\
\hline
$0.04$              & $0.77621$     & $1.21415$      & $4.74675\times 10^{-6}$ &$8.2565\times10^{-2}$& $1.5360\times10^{-13}$\\
\hline
$0.05$              & $0.77643$     & $1.21242$      & $4.73825\times 10^{-6}$ &$8.2439\times10^{-2}$& $1.5258\times10^{-13}$\\
\hline
$0.06$              & $0.77671$     & $1.21074$      & $4.72994\times 10^{-6}$ &$8.2313\times10^{-2}$& $1.5158\times10^{-13}$\\
\hline
$0.07$              & $0.77703$     & $1.20913$      & $4.72182\times 10^{-6}$ &$8.2188\times10^{-2}$& $1.5060\times10^{-13}$\\
\hline
$0.08$              & $0.77740$     & $1.20757$      & $4.71390\times 10^{-6}$ &$8.2063\times10^{-2}$& $1.4964\times10^{-13}$\\
\hline
$0.09$              & $0.77782$     & $1.20607$      & $4.70618\times 10^{-6}$ &$8.1939\times10^{-2}$& $1.4870\times10^{-13}$\\
\hline
$0.10$              & $0.77829$     & $1.20463$      & $4.69867\times 10^{-6}$ &$8.1815\times10^{-2}$& $1.4778\times10^{-13}$\\
\hline
\end{tabular}
\end{center}
\caption{Masses of $\r$-meson, $a_1$-meson and pion. $f_{\pi}$ is
the pion decay constant. $\Delta$ is the deviation from GOR
relation. $m_q=0$, $z_{IR}=1/(0.3227\mathrm{GeV})$ and
$\sigma_0=(0.304\mathrm{GeV})^3$.}\label{tab:mass1}
\end{table}

\subsection{Binding energy of heavy quarkonium}\label{sec:breaking1}
We study the binding energy of heavy quarkonium in the confining
phase where the quark density is (\ref{eq:con_den}) and the
asymptotic geometry is (\ref{eq:bgdzeromq}) with $M=0$. We observe
breaking of a fundamental string which connects two heavy quarks. It
describes dissociation of a bound state of heavy quarks into two
heavy-light quark bound states. The action for a fundamental string
is
\begin{eqnarray}\label{eq:NGact0}
S=\frac{1}{2\pi\alpha^\prime}\int
d^2\sigma\sqrt{\det\partial_aX^M\partial_bX^NG_{MN}},
\end{eqnarray}
where $a,\,b$ are the string worldsheet indices and $G_{MN}$ is the
metric (\ref{eq:metricansatz}). We take $R=1$ and
$\alpha^\prime=1/2\pi$. We choose a gauge condition and an ansatz
for the coordinate $z$ as
\begin{eqnarray}
\tau=t,~ \sigma_1=x^1\equiv x~\mathrm{and}~z=z(x).
\end{eqnarray}
The action becomes
\begin{eqnarray}\label{eq:NGact}
S=\int^T_0dt\int^{r/2}_{-r/2}dx\frac{1}{z^2}\sqrt{F(G+z^{\prime2})}
.
\end{eqnarray}
Notice that we use the Fefferman-Graham coordinate and $F$ and $G$
are solutions obtained by numerically solving \eq{eq:geoeom}, whose
asymptotic expansions at the boundary are given in
\eq{eq:bgdzeromq}. The Hamiltonian density is
\begin{eqnarray}\label{eq:ham1}
{\mathcal{H}}=-\frac{1}{z^2}\frac{FG}{\sqrt{F(G+z^{\prime2})}}.
\end{eqnarray}
We define $z=z_0$ as the point where $\left.\frac{\partial
z}{\partial x}\right|_{z=z_0}=0$. The Hamiltonian density at $z=z_0$
becomes
\begin{eqnarray}\label{eq:ham2}
{\mathcal{H}}_0=-\frac{1}{z_0^2}\sqrt{F_0G_0},
\end{eqnarray}
where $F_0$ and $G_0$ are the values of $F$ and $G$ at $z=z_0$. As
Hamiltonian is conserved we can get the relation between $r$ and
$z_0$ from (\ref{eq:ham1}) and (\ref{eq:ham2}) as
\begin{eqnarray}\label{eq:dis}
r=2\int^{z_0}_0dz\frac{z^2\sqrt{F_0G_0}}{\sqrt{z_0^4FG^2-z^4F_0G_0G}}.
\end{eqnarray}
The kinetic energy of two heavy quarks can be ignored since we are
considering a static configuration. The potential energy is obtained
from (\ref{eq:NGact}) and (\ref{eq:dis}) as
\begin{eqnarray}
V=2\int^{z_0}_0dz\frac{z_0^2}{z^2}\frac{FG}{\sqrt{z_0^4FG^2-z^4F_0G_0G}}.
\end{eqnarray}
The potential energy diverges since the heavy quarks have infinite
masses at the boundary. We regularize the potential energy by
subtracting the energy of two straight strings corresponding to two
free heavy quarks. By choosing a gauge condition and an ansatz for
the two strings as
\begin{eqnarray}
\tau=t,~ \sigma_1=z~\mathrm{and}~x=\mathrm{constant},
\end{eqnarray}
the energy of two static straight strings is
\begin{eqnarray}
V_0=2\int^{z_{\sss{IR}}}_0\frac{1}{z^2}\sqrt{F}.
\end{eqnarray}
The regularized binding energy of heavy quarkonium is
\begin{eqnarray}
E=2\int^{z_0}_0dz\frac{z_0^2}{z^2}\frac{FG}{\sqrt{z_0^4FG^2-z^4F_0G_0G}}
-2\int^{z_{\sss{IR}}}_0\frac{\sqrt{F}}{z^2}.\label{eq:heavypot}
\end{eqnarray}
The dissociation length increases with the quark density as shown in
Table \ref{tab:sb1}. It indicates that it takes more energy to
produce a pair of heavy-light quark bound states as the quark
density increases. By taking account of (\ref{eq:con_den}), this is
consistent with the result of \cite{Lee:2010dh} that the
dissociation length increases as the chiral condensate decreases.

\begin{table}[h!]
\begin{center}
\begin{tabular}{|c|c|}
\hline
$Q(\mathrm{GeV}^3)$ & length with $m_q=0 (\mathrm{GeV}^{-1})$    \\
\hline \hline
$0.00$          & $2.2217$    \\
\hline
$0.01$          & $2.2218$    \\
\hline
$0.02$          & $2.2220$    \\
\hline
$0.03$          & $2.2224$    \\
\hline
$0.04$          & $2.2228$    \\
\hline
$0.05$          & $2.2234$    \\
\hline
$0.06$          & $2.2242$    \\
\hline
$0.07$          & $2.2251$    \\
\hline
$0.08$          & $2.2261$    \\
\hline
$0.09$          & $2.2272$    \\
\hline
$0.10$          & $2.2285$    \\
\hline
\end{tabular}
\end{center}
\caption{Dissociation length. $m_q=0$,
$z_{IR}=1/(0.3227\mathrm{GeV})$ and
$\sigma_0=(0.304\mathrm{GeV})^3$.}\label{tab:sb1}
\end{table}

\section{Mesons with a nonzero $m_q$}
\subsection{Light meson spectra}
\begin{figure}[h!]
\begin{center}
\epsfxsize=7cm
   \epsfbox{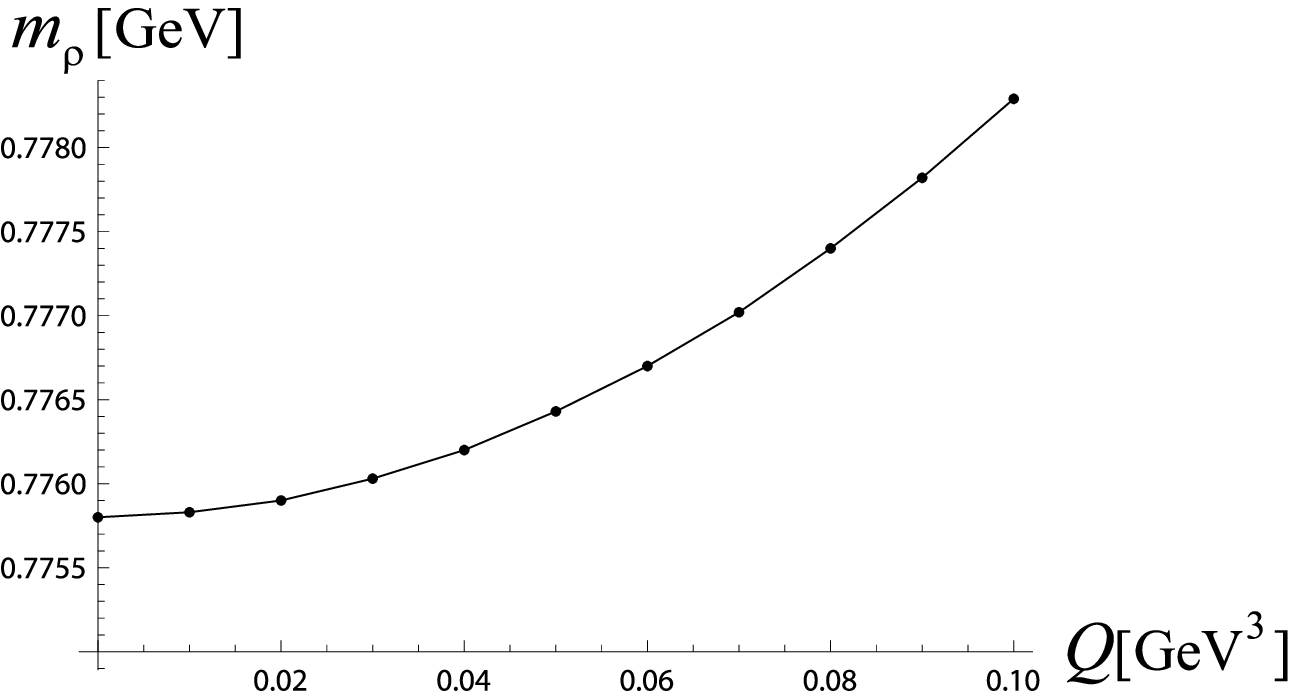}\\
   (a)\\~\\
$\begin{array}{cc}
  \epsfxsize=7cm
   \epsfbox{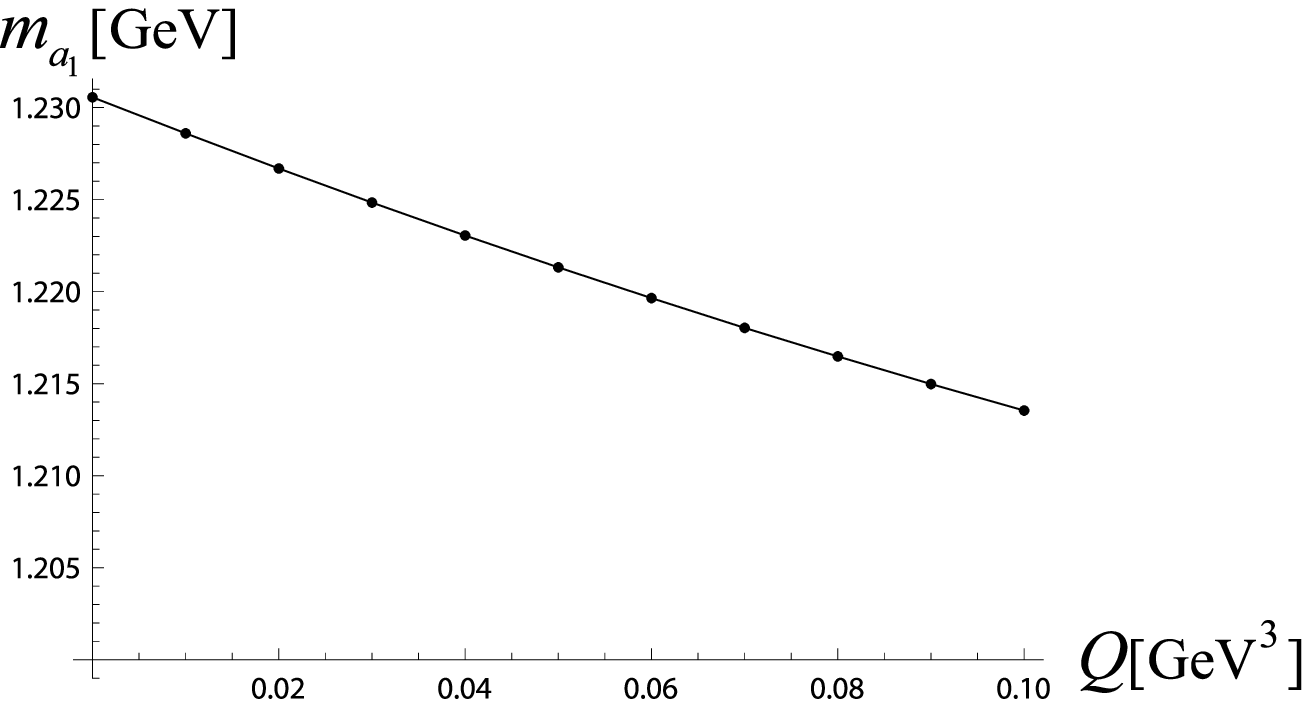}
  &
  \epsfxsize=7cm
   \epsfbox{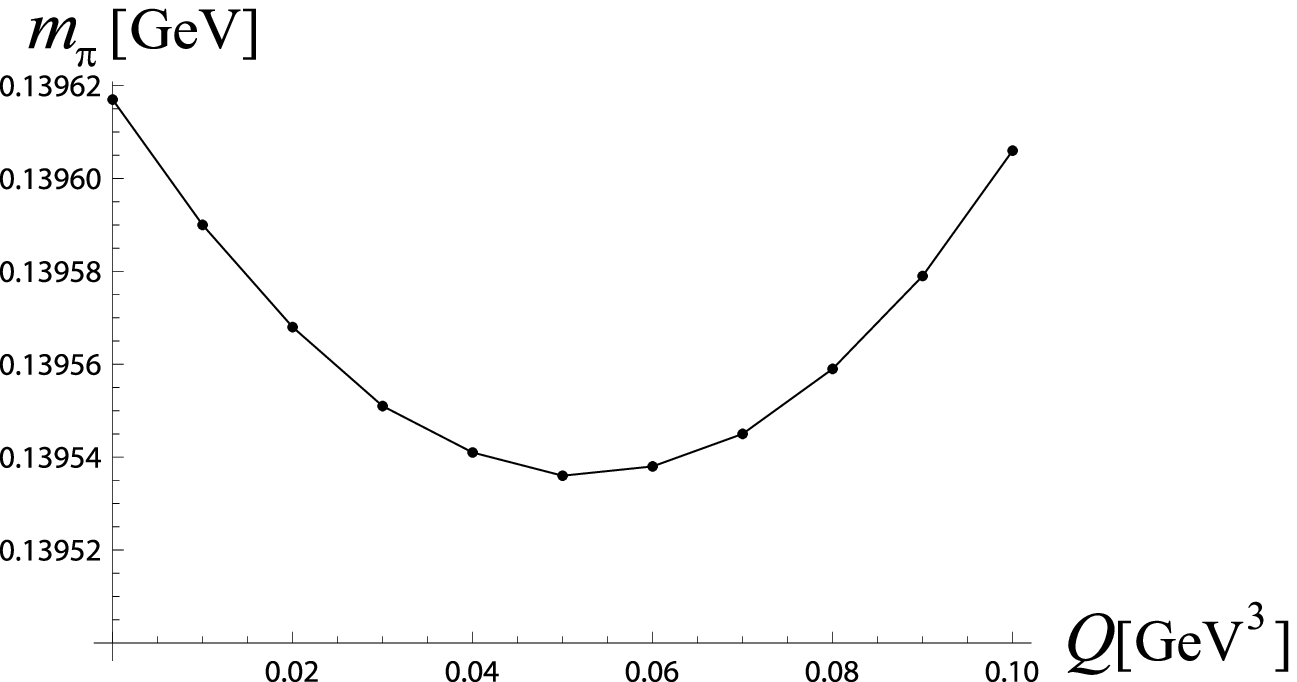}\\
   \mathrm{(b)}&\mathrm{(c)}
\end{array}$
  \caption{ Meson masses with $m_q=0.002383\mathrm{GeV}$. (a)$\r$-meson
   (b)$a_1$-meson
   (c)pion}\label{fig:mass2}
 \end{center}
\end{figure}

\begin{figure}[h!]
\begin{center}
$\begin{array}{cc} \epsfxsize=7cm \epsfbox{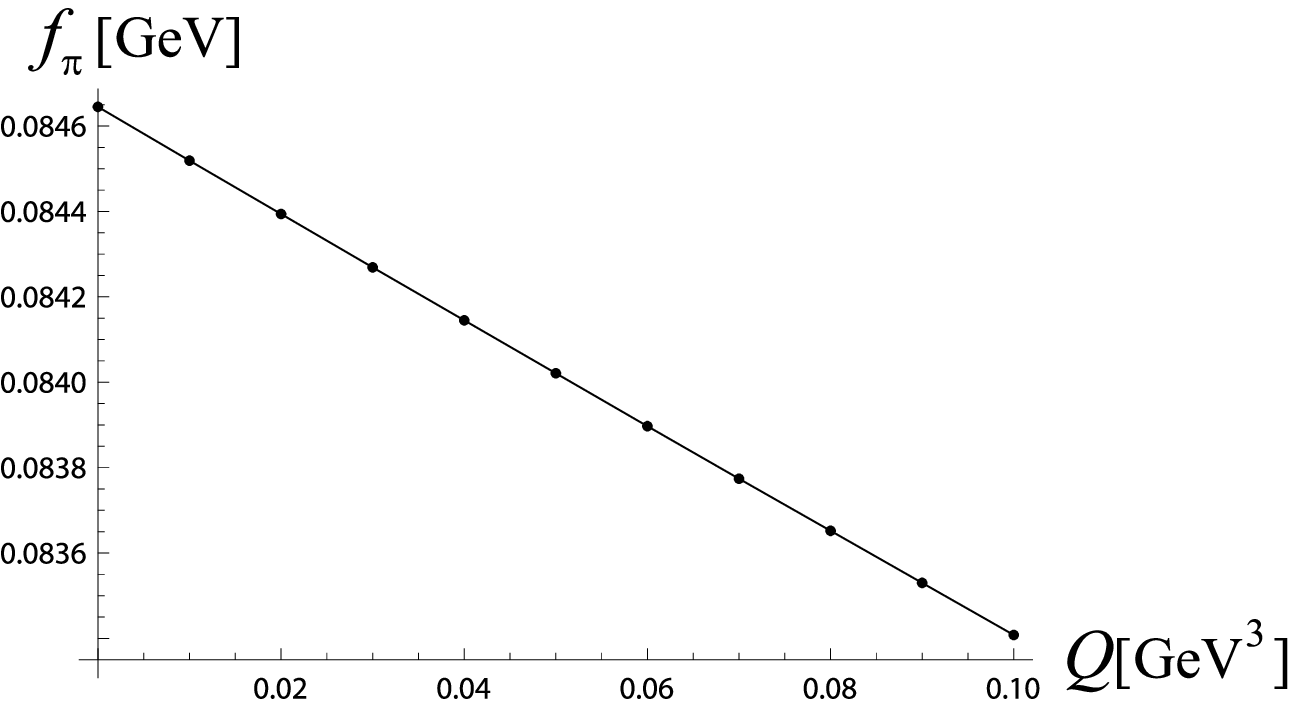} &\epsfxsize=7cm
\epsfbox{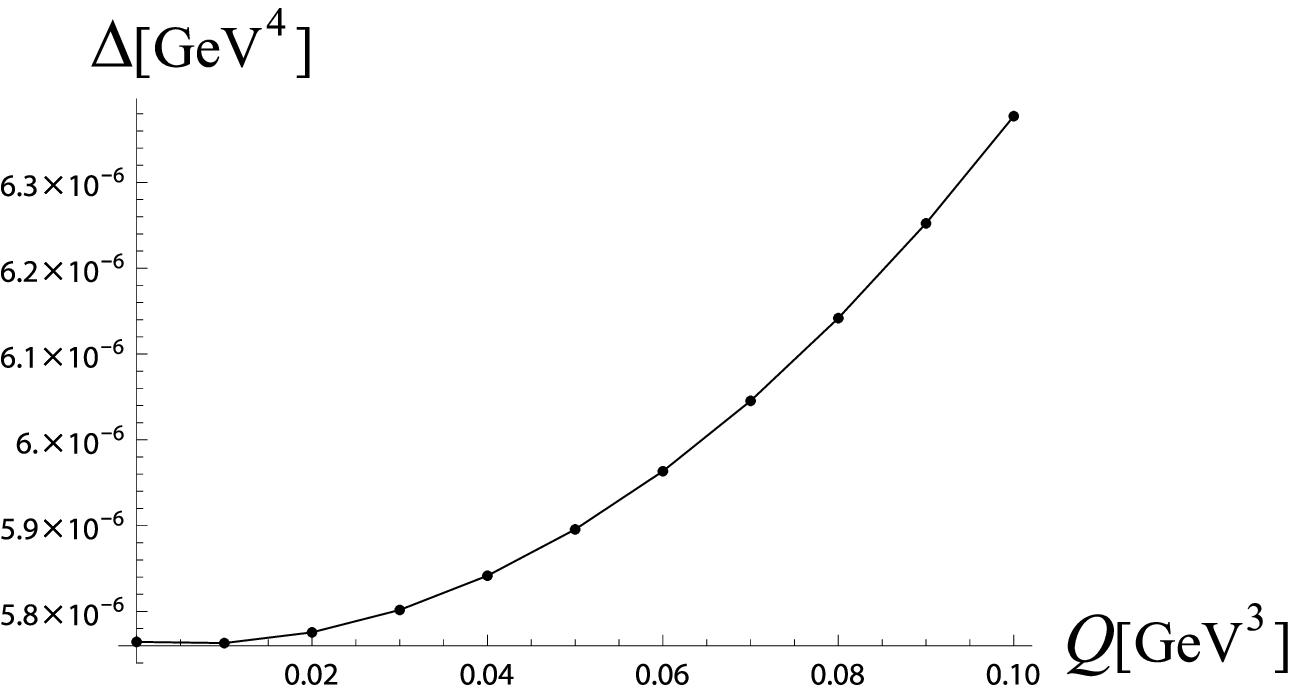}
\end{array}$
\caption{(a)pion decay constant (b)deviation from GOR relation.
$m_q=0.002383\mathrm{GeV}$.}\label{fig:decay_gor2}
\end{center}
\end{figure}

We solve (\ref{eq:geoeom}) with a nonzero quark mass,
$m_q=0.002383\mathrm{GeV}$ \cite{Lee:2010dh}, which breaks the
chiral symmetry explicitly. The asymptotic solutions are
\begin{eqnarray}
F(z)&=&1-\frac{1}{12}\kappa^2m_q^2
z^2+\frac{1}{144}(\kappa^4m_q^4-18\kappa^2m_q\sigma-3\kappa^4m_q^4\log
z)z^4\nonumber\\
&&+\frac{1}{31104}\Big(17280\frac{\kappa^2Q^2}{g^2}+65\kappa^6m_q^6-396\kappa^4m_q^3\sigma-2592\kappa^2\sigma^2\nonumber\\
&&~-66\kappa^6m_q^6\log z-864\kappa^4m_q^3\sigma\log
z-72\kappa^6m_q^6(\log z)^2\Big)z^6+\cdots\nonumber\\
G(z)&=&1-\frac{1}{12}\kappa^2m_q^2z^2+\frac{1}{144}(\kappa^4m_q^4-18\kappa^2m_q\sigma-3\kappa^4m_q^4\log
z)z^4\nonumber\\
&&+\frac{1}{31104}\Big(-3456\frac{\kappa^2Q^2}{g^2}+65\kappa^6m_q^6-396\kappa^4m_q^3\sigma-2592\kappa^2\sigma^2\nonumber\\
&&~-66\kappa^6m_q^6\log z-864\kappa^4m_q^3\sigma\log
z-72\kappa^6m_q^6(\log z)^2\Big)z^6+\cdots\nonumber\\
\phi(z)&=&m_qz+\Big(\sigma+\frac{1}{6}\kappa^2m_q^3\log
z\Big)z^3+\frac{1}{576}(-13\kappa^4m_q^5+144\kappa^2m_q^2\sigma+24\kappa^4m_q^5\log
z)z^5\nonumber\\
&&+\frac{1}{62208}\Big(-1728\kappa^2\frac{m_qQ^2}{g^2}-209\kappa^6m_q^7
+1044\kappa^4m_q^4\sigma+10368\kappa^2m_q\sigma^2\nonumber\\
&&~+174\kappa^6m_q^7\log z+3456\kappa^4m_q^4\sigma\log
z+288\kappa^6m_q^7(\log z)^2\Big)z^7+\cdots\nonumber\\
A_0(z)&=&\mu-Qz^2-\frac{1}{24}\kappa^2m_q^2Qz^4+\frac{1}{864}(Q\kappa^4m_q^4-36Q\kappa^2m_q\sigma-6Q\kappa^4m_q^4\log
z)z^6\nonumber\\
&&-\frac{1}{248832}\Big(27648\frac{Q^3\kappa^2}{g^2}-205Q\kappa^6m_q^6+1872Q\kappa^4m_q^3\sigma+5184Q\kappa^2\sigma^2\nonumber\\
&&~+312Q\kappa^6m_q^6\log z+1728Q\kappa^4m_q^3\sigma\log
z+144Q\kappa^6m_q^6(\log z)^2\Big)z^8+\cdots\label{eq:bgdnonzeromq}.
\end{eqnarray}
To investigate the meson masses, we solve (\ref{eq:mesonmass}) on
the full numerical solutions obtained from (\ref{eq:bgdnonzeromq}).
We also use the relation (\ref{eq:con_den}) with $Q_0$ as a unit and
the values of the chiral condensate and IR cutoff (\ref{eq:sig_ir}).
The result is shown in Table \ref{tab:mass2}, Figure \ref{fig:mass2}
and Figure \ref{fig:decay_gor2}.

\begin{table}[h!]
\begin{center}
\begin{tabular}{|c|c|c|c|c|c|}
\hline
$Q(\mathrm{GeV}^3)$ &$m_{\rho}(\mathrm{GeV})$ & $m_{a_1}(\mathrm{GeV})$ & $m_\pi(\mathrm{GeV})$ &$f_\pi(\mathrm{GeV})$ &$\Delta(\mathrm{GeV}^4)$         \\
\hline \hline
$0.00$              & $0.77580$     & $1.23056$      & $0.139617$  & $8.4645\times10^{-2}$ & $5.7645\times10^{-6}$ \\
\hline
$0.01$              & $0.77583$     & $1.22860$      & $0.139590$  & $8.4519\times10^{-2}$ & $5.7631\times10^{-6}$ \\
\hline
$0.02$              & $0.77590$     & $1.22669$      & $0.139568$  & $8.4394\times10^{-2}$ & $5.7756\times10^{-6}$ \\
\hline
$0.03$              & $0.77603$     & $1.22484$      & $0.139551$  & $8.4269\times10^{-2}$ & $5.8017\times10^{-6}$ \\
\hline
$0.04$              & $0.77620$     & $1.22305$      & $0.139541$  & $8.4145\times10^{-2}$ & $5.8416\times10^{-6}$ \\
\hline
$0.05$              & $0.77643$     & $1.22132$      & $0.139536$  & $8.4021\times10^{-2}$ & $5.8956\times10^{-6}$ \\
\hline
$0.06$              & $0.77670$     & $1.21965$      & $0.139538$  & $8.3897\times10^{-2}$ & $5.9634\times10^{-6}$ \\
\hline
$0.07$              & $0.77702$     & $1.21803$      & $0.139545$  & $8.3774\times10^{-2}$ & $6.0454\times10^{-6}$ \\
\hline
$0.08$              & $0.77740$     & $1.21648$      & $0.139559$  & $8.3652\times10^{-2}$ & $6.1418\times10^{-6}$ \\
\hline
$0.09$              & $0.77782$     & $1.21498$      & $0.139579$  & $8.3530\times10^{-2}$ & $6.2523\times10^{-6}$ \\
\hline
$0.10$              & $0.77829$     & $1.21354$      & $0.139606$  & $8.3408\times10^{-2}$ & $6.3772\times10^{-6}$ \\
\hline
\end{tabular}
\end{center}
\caption{Masses of $\r$-meson, $a_1$-meson and pion. $f_{\pi}$ is
the pion decay constant. $\Delta$ is the deviation from GOR
relation. $m_q=0.002383\mathrm{GeV}$,
$z_{IR}=1/(0.3227\mathrm{GeV})$ and
$\sigma_0=(0.304\mathrm{GeV})^3$.}\label{tab:mass2}
\end{table}

The $\r$-meson mass increases whereas the $a_1$-meson mass decreases
as the quark density increases. The patterns are the same as the
results with a zero quark mass as observed in Section
\ref{sec:masszeromq}. The $a_1$-meson mass decreases with the quark
density in both cases of zero and nonzero quark masses. By taking
the relation (\ref{eq:con_den}) into account, the spectrum is
consistent with the result of \cite{Lee:2010dh} that the $a_1$-meson
mass increases with the chiral condensate, but contrary to the
result of \cite{Jo:2009xr} that the $a_1$-meson mass increases with
the quark density. It shows that the effect of the chiral condensate
on the mass spectra dominates the effect of the quark density. A
distinguishing feature is that the spectrum of pion exhibits a
critical point in the case of a nonzero quark mass whereas it
decreases monotonically with the quark density in the case of a zero
quark mass as observed in Section \ref{sec:masszeromq}. At the low
quark density, the chiral condensate is high so that the quark mass
is negligible. Thus the pion mass spectrum is similar to the case of
a zero quark mass. At the high quark density, however, the chiral
condensate is small. The effect of the quark mass becomes comparable
with the chiral condensate near the critical point. Above the
critical point, the effect of the quark mass, which is constant,
dominates the chiral condensate so that the quark density mainly
contributes to the medium. The pion mass therefore increases with
the quark density as it is observed in \cite{Jo:2009xr}.

The pion decay constant decreases as the quark density increases.
The deviation from the GOR relation (\ref{eq:gor}) is
$\Delta=f_\pi^2m_\pi^2-2m_q\sigma$. The GOR relation is satisfied up
to $10^{-6}\mathrm{GeV}^4$.

\subsection{Binding energy of heavy quarkonium}
We study the binding energy of heavy quarkonium in the confining
phase. We numerically solve the equations (\ref{eq:geoeom}) with
(\ref{eq:bgdnonzeromq}) as asymptotic expansions of the metric
components at the boundary. The binding energy of heavy quarkonium
is (\ref{eq:heavypot}) and the dissociation occurs when $E=2m_q$.
The result is shown in Table \ref{tab:sb2}.

The quark mass is small so that there is no big difference between
the cases of zero and nonzero quark masses as shown in Table
\ref{tab:sb1} and Table \ref{tab:sb2}. The dissociation length with
a nonzero quark mass is slightly longer. It implies that the quark
mass shifts the binding energy a little bit higher.

\begin{table}[h!]
\begin{center}
\begin{tabular}{|c|c|}
\hline
$Q(\mathrm{GeV}^3)$  & length with $m_q\neq0 (\mathrm{GeV}^{-1})$   \\
\hline \hline
$0.00$          & $2.2382$  \\
\hline
$0.01$          & $2.2383$  \\
\hline
$0.02$          & $2.2385$  \\
\hline
$0.03$          & $2.2389$  \\
\hline
$0.04$          & $2.2394$  \\
\hline
$0.05$          & $2.2400$  \\
\hline
$0.06$          & $2.2407$  \\
\hline
$0.07$          & $2.2416$  \\
\hline
$0.08$          & $2.2426$  \\
\hline
$0.09$          & $2.2438$  \\
\hline
$0.10$          & $2.2450$  \\
\hline
\end{tabular}
\end{center}
\caption{Dissociation length. $m_q=0.002383\mathrm{GeV}$,
$z_{IR}=1/(0.3227\mathrm{GeV})$ and
$\sigma_0=(0.304\mathrm{GeV})^3$.}\label{tab:sb2}
\end{table}

\section{Discussion}
We have studied subleading $1/N_c$ corrections from the chiral
condensate and the quark density to the meson spectra and the
binding energy of heavy quarkonium. We have considered the
gravitational backreaction of a massive scalar field, which
corresponds to the current quark mass and the chiral condensate, and
the time component of $U(1)$ gauge fields, which corresponds to the
chemical potential and the quark density. The geometries are
numerically solved, separately with zero and nonzero current quark
masses. The geometries are constrained by a model-independent
relation in QCD phenomenology that the chiral condensate gets
reduced linearly with the quark density. The meson masses and
binding energy of heavy quarkonium are calculated on each geometry.

With a zero quark mass, the $\r$-meson mass increases as the quark
density increases. This is consistent with the results of
\cite{Jo:2009xr,Lee:2010dh}, where the meson spectra are studied
with the quark density and the chiral condensate as parameters
separately. The $a_1$-meson mass and pion mass decrease with the
quark density. The results are consistent with the spectra with the
chiral condensate, but contrary to the spectra with the quark
density.

With a nonzero quark mass, the spectra of the $\r$-meson and
$a_1$-meson present the same pattern as the spectra with a zero
quark mass. The $a_1$-meson mass decreases with the quark density in
both cases of zero and nonzero quark masses. By taking account of
the fact that the chiral condensate gets reduced with the quark
density, this is consistent with the result of \cite{Lee:2010dh}
that the $a_1$-meson mass increases with the chiral condensate, but
contrary to the result of \cite{Jo:2009xr} that $a_1$-meson mass
increases with the quark density. It indicates that the effect of
the chiral condensate on the mass spectra dominates the effect of
the quark density. The mass spectrum of pion exhibits a critical
point. Below the critical point, where the quark density is low, the
chiral condensate is high so the quark mass is negligible. Thus the
pion mass decreases with the quark density as it does with a zero
quark mass. Above the critical point, where the quark density is
high and the chiral condensate is low, the effect of the constant
quark mass dominates the chiral condensate. As the quark density
becomes the main contribution to the medium, the pion mass increases
with the quark density as observed in \cite{Jo:2009xr}.

The dissociation length of heavy quarkonium increases with the quark
density for both cases of zero and nonzero quark masses. It shows
that it requires more energy to produce a pair of heavy-light quark
bound states as the quark density increases. This result is
consistent with the dissociation length depending on the chiral
condensate \cite{Lee:2010dh}. The dissociation length is slightly
longer with a nonzero quark mass. It indicates that the quark mass
shifts the binding energy higher.

\vspace{1cm}

{\bf Acknowledgement}

This work was supported by the National Research Foundation of Korea(NRF) grant funded by
the Korea government(MEST) through the Center for Quantum Spacetime(CQUeST) of Sogang
University with grant number 2005-0049409. C. Park was also
supported by Basic Science Research Program through the
National Research Foundation of Korea(NRF) funded by the Ministry of
Education, Science and Technology(2010-0022369).

\end{document}